\documentclass[useAMS,usenatbib]{mn2e}
\usepackage{graphicx}
\title[IMBH in AGN disks I.]{Intermediate mass black holes in AGN disks I. Production \& Growth}

\author[B. McKernan, K.E.S. Ford, W.Lyra \& H.B. Perets]{B. McKernan$^{1,2,3}$\thanks{E-mail:bmckernan at amnh.org (BMcK)}, K.E.S. Ford$^{1,2,3}$, W.Lyra$^{2,4,6}$ \& H.B. Perets $^{5}$ \\
$^{1}$Department of Science, Borough of Manhattan Community College, City University of New York, New York, NY 10007, USA\\
$^{2}$Department of Astrophysics, American Museum of Natural History, New York, NY 10024, USA\\
$^{3}$Graduate Center, City University of New York, 365 5th Avenue, New York, NY 10016, USA\\
$^{4}$Jet Propulsion Laboratory,California Institute of Technology, 4800 Oak Grove Drive, Pasadena, CA 91109, USA\\
$^{5}$Harvard-Smithsonian Center for Astrophysics, 60 Garden St., Cambridge, MA 02138, USA\\
$^{6}$NASA Carl Sagan Fellow\\
\\
}
\begin{document}

\date{Accepted. Received; in original form}

\pagerange{\pageref{firstpage}--\pageref{lastpage}} \pubyear{2008}

\maketitle

\label{firstpage}

\begin{abstract}
Here we propose a mechanism for efficiently growing intermediate mass black 
holes (IMBH) in disks around supermassive black holes. Stellar mass objects can 
efficiently agglomerate when facilitated by the gas disk. Stars, compact objects 
and binaries can migrate, accrete and merge within disks around supermassive 
black holes. While dynamical heating by cusp stars excites the 
velocity dispersion of  nuclear cluster objects (NCOs) in the disk, gas in the 
disk damps NCO orbits. If gas damping dominates, NCOs remain in the disk 
with circularized orbits and large collision cross-sections. IMBH seeds can 
grow extremely rapidly by collisions with disk NCOs 
at low relative velocities, allowing for super-Eddington growth rates. Once an 
IMBH seed has cleared out its feeding zone of disk NCOs, 
growth of IMBH seeds can become dominated by gas accretion from the 
AGN disk. However, the IMBH can migrate in the disk and expand its feeding zone, 
permitting a super-Eddington accretion rate to continue. Growth of IMBH seeds via
 NCO collisions is enhanced by a pile-up of migrators. 

We highlight the 
remarkable parallel between the growth of IMBH in AGN disks with models of giant 
planet growth in protoplanetary disks. If an IMBH becomes massive enough it can 
open a gap in the AGN disk. IMBH migration in AGN disks may stall, allowing them 
to survive the end of the AGN phase and remain in galactic nuclei. Our proposed 
mechanisms should be more efficient at growing IMBH in AGN disks than the 
standard model of IMBH growth in stellar clusters. Dynamical heating of disk NCOs
 by cusp stars is transferred to the gas in a AGN disk helping to maintain the outer disk 
against gravitational instability. Model predictions, observational constraints and 
implications are discussed in a companion paper (Paper II).
\end{abstract}


\section{Introduction}
\label{sec:intro}
Extensive evidence exists that supermassive black holes ($>10^{6}M_{\odot}$) 
are found in the centers of most galaxies \citep[e.g.][]{b5}. Extensive evidence
 also exists for stellar mass black holes in our own Galaxy \citep{b68}. Stellar mass black holes 
are expected to form as the end product of high-mass stars. 
Supermassive black holes, by contrast, have grown to their current size over cosmic time, from 
much smaller seeds \citep[e.g.][ \& references therein]{b18,b16,b15,b90}. 
Intermediate mass black 
holes (IMBH; $\sim 10^{2}-10^{4}M_{\odot}$) may have been the original seeds for 
supermassive black holes or, they may have contributed to fast early growth of 
such seeds via mergers \citep[e.g.][]{b86,b87}. Though we expect IMBH should 
exist, at least as an intermediate stage on the way to a supermassive black hole,
observationally the evidence for their existence is scant and ambiguous, 
especially compared with evidence for supermassive and stellar mass black holes. 
The low mass end of the supermassive black hole distribution in galactic nuclei 
may extend down to $\sim 10^{5}M_{\odot}$ \citep{b73}, but below this mass the 
evidence becomes ambiguous. The ultra-luminous X-ray sources 
(ULXs) observed outside galactic nuclei \citep[e.g.][]{b54} may be powered by 
accretion onto IMBH \citep{b87}. However ULXs could 
also be a explained by beamed radiation from accreting stellar-mass black holes 
\citep{b82} and power-law dominated ULXs might be due to background AGN. IMBH 
have so far been hard to find and constrain in the local Universe, either in our 
own Galaxy or at low z.

Active galactic nuclei (AGN) are believed to be powered by accretion onto a 
supermassive black hole. The accretion disk should contain a population of 
stars and compact objects (collectively nuclear cluster objects, NCOs) that can 
migrate within and accrete from the disk 
\citep[e.g.][]{b57,b58,b51,b44,b45,b35,b96,b99}. In \citet{b96} we speculated 
that IMBH seeds may form efficiently in AGN disks due to NCO collisions and 
mergers, which is quite different from the standard model of stellar mass black 
holes merging in stellar clusters \citep[e.g.][]{b81,b87}. Here we argue that 
IMBH production is in fact far more likely and more efficient in AGN disks, with 
implications for AGN observations, duty cycle and supermassive black hole 
accretion rates.

In this paper (and its companion, Paper II, McKernan et al. 2012) we discuss 
semi-analytically the production of intermediate mass black holes in the 
environment of AGN disks. Discussion of observational predictions of this 
model of IMBH growth as well as 
consequences for AGN disks, duty cycles and the demographics of activity in 
galactic nuclei at low and high redshift will be left to Paper II. In section 
\S\ref{sec:collisions}, we discuss 
why we think IMBH can be built in AGN disks. In section \S\ref{sec:grow} we 
explore mechanisms that will be important in actually growing IMBH in AGN disks, 
including the competing forces of eccentricity damping and excitation in the 
disk. The importance of IMBH migration is outlined in section \S\ref{sec:mig}. 
Section~\S\ref{sec:model_growth} outlines a simple model of IMBH growth in AGN 
disks and we highlight the remarkable parallel between the growth of IMBH in AGN 
disks and the growth of giant planets in protoplanetary disks. Finally in section
 \S\ref{sec:conclusions}, we outline our conclusions and future work.

\section{Why IMBH can be built in AGN disks}
\label{sec:collisions}
The largest, supermassive, black holes in the Universe
($M_{\rm BH}\sim 10^{6}-10^{9}M_{\odot}$) live in galactic centers
\citep[e.g.][]{b5}. We expect a dense nuclear cluster of objects to surround 
the supermassive black hole as a result of stellar evolution, dynamical
friction, secular evolution and minor mergers \citep[e.g.][]{b12,b11,b8}. In 
our own Galaxy, the distributed mass within $\sim 1$pc of Sgr A* is 
$\sim 10-30\%$ of the mass of the supermassive black hole \citep{b17}. If a 
large quantity of gas somehow arrives in the innermost pc of a galactic nucleus 
\citep[e.g.][]{b39,b92,b93,b94,b95},
it will likely lose angular momentum and accrete onto the central supermassive
 black hole. But in doing so, gas must also interact with the NCO population.
Depending on the aspect ratio of the disk that forms, a few percent
of NCO orbits are likely to coincide with the accretion flow. A small
percentage of NCO orbits coincident with the geometric cross-section of a thin 
disk would lead to an initial population of  $\sim 10^{3}M_{\odot}$ of NCOs in 
a pc-scale accretion flow around a SgrA* sized black hole.

NCOs can exchange angular momentum with gas in the disk, and each other, so 
they can scatter each other and migrate within the disk \citep[see][]{b96}. 
The processes involved are analagous to protoplanetary disk theory 
\citep[e.g.][]{b91,b20}. Indeed, physical conditions in the \emph{outskirts} of 
AGN disks are relatively close to those in protoplanetary disks \citep{b96}. The 
migration of NCOs in the disk will enhance the probability of collisions, 
mergers and ejections. Under these conditions, IMBH seeds can grow. IMBH seeds 
will be objects $\geq 10M_{\odot}$ that will not lose very much mass (e.g. 
stellar mass black holes, hard massive binaries). IMBH 'seedlings' we define as 
objects $ \geq 10M_{\odot}$ that have grown via mergers (e.g. 
the merged end-product of a hard binary). IMBH seeds and seedlings located at 
semi-major axis $a$ in an AGN disk will maintain a 'feeding zone' within which 
they may collide with nearly 
co-orbital disk NCOs. By analogy with proto-planet growth we define the feeding 
zone to be $a \pm 4R_{H}$ where $a$ is the IMBH semi-major axis and 
$R_{H}=a(q/3)^{1/3}$ is the IMBH Hill radius, with $q$ the mass ratio of 
IMBH:supermassive black hole. Once the object gets to $\geq 100M_{\odot}$ we 
will call the result an IMBH. 

The disk NCO population is subjected to dynamical heating from cusp 
stars and dynamical cooling from gas damping. If gas damping dominates, IMBH 
seeds and disk
 NCOs will have their orbits rapidly damped. As a result their collision 
cross-sections will rapidly increase (since the relative velocity of encounters 
will be small, particularly in the outer disk). IMBH seedlings will initially 
accrete disk NCOs within their 
feeding zone in a 'core accretion' mode of growth. Once nearby NCOs have been 
scattered or accreted, gas accretion dominates IMBH growth. However, the 
migration of IMBH seedlings within the disk allows growth to continue via 
collisions as well as via gas accretion. Thus, we expect IMBH to grow 
within AGN disks, analagous to the growth of giant planets within protoplanetary 
disks and we expect the IMBH growth rate will be much larger than in stellar 
clusters. 

Here we concentrate on growing IMBH within the AGN disk itself. Of course it is 
possible that IMBH already exist in the galactic nucleus when 
the AGN disk first forms. A top-heavy initial mass function of cusp stars can 
lead to IMBH seedling formation before low angular momentum gas arrives in 
the nucleus. IMBH can also arrive from outside the galactic nucleus to interact 
with the AGN disk since mass segregation and dynamical friction can deliver IMBH 
to the central parsec of galactic nuclei in a few Gyrs from nearby clusters 
\citep[e.g.][see also Paper II]{b99}. 

The physics involved in IMBH formation in AGN disks spans multiple regimes and 
physical processes and would usefully benefit from detailed numerical 
simulations. Such simulations require 
realistic treatments of (amongst others): N-body collisions, mergers, accretion, 
tidal forces, gravitational radiation, Special \& General Relativity, radiative 
transfer, the magneto-rotational instability and the gravitational instability. 
However at present there are no simulations that can adequately 
address the relevant physics in a self-consistent manner. We take a semi-analytic
 approach following  \citep[e.g.][]{b81} on the build-up of IMBH in star 
clusters, \citep[e.g.][]{b34} on stellar dynamical heating and cooling, as well 
as formalism on planet growth from protoplanetary theory \citep[e.g.][\& 
references therein]{b91,b20}. 
 
\section{How to build IMBH in AGN disks}
\label{sec:grow}
In this section, we shall outline the key phenomena involved in growing IMBH 
seeds in AGN disks. In order to grow into IMBH, seeds must collide with and 
accrete mass, either NCOs or gas. In section 
\S\ref{sec:ecc} we discuss NCO collision cross sections in AGN disks and the 
importance of eccentricity damping and excitation. In section~\S\ref{sec:damp} we 
outline a model of dynamical heating and cooling of NCO orbits in AGN disks and 
we discuss the implications for IMBH seed growth. Section \S\ref{sec:bins}
briefly outlines issues involved in merging binaries, a potentially important 
channel for producing IMBH seedlings. 

\subsection{Collision cross-sections in the disk}
\label{sec:ecc}
\begin{figure}
\includegraphics[width=3.35in,height=3.35in,angle=-90]{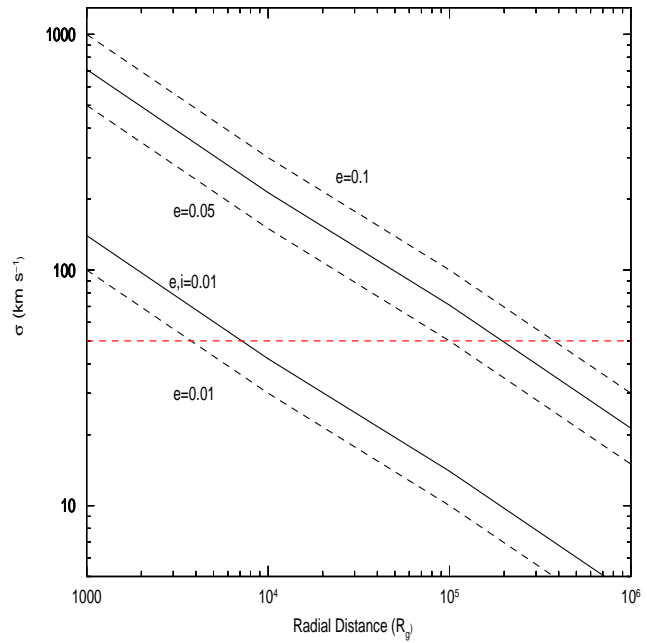}
\caption{The velocity dispersion ($\sigma \approx \sqrt{e^{2}+i^{2}}V_{kep}$) of 
NCOs in a Keplerian AGN disk. Shown are $\sigma$ as a function of disk radius, 
for eccentricity and inclination values of (e,i)=0.01,0.05 (solid lines) and 
$e=0.01,0.05,0.1$, with $i=0$(dashed lines). Also shown (red dashed horizontal 
line) is the typical velocity dispersion in star clusters ($\sim 50$ km 
$\rm{s}^{-1}$). Note that most disk NCOs should live in the outer disk 
($>10^{4}r_{g}$) for an NCO population that grows as $r^{2}$.
\label{fig:relvel}}
\end{figure}

In the absence of disk gas, NCOs change their orbits only due to weak 
gravitational interactions, occuring on the (long) relaxation timescale (see 
below). The interaction with a gaseous disk gives rise to new effects, namely 
torques. NCOs embedded in the disk, or crossing it,
will have their eccentricities and inclinations damped relative to the disk. Such 
processes should enhance the stellar density in the disk region and lower the 
velocity dispersion of NCOs embedded in the disk, in the absence of other 
important effects. This, in turn, gives rise to a 
higher rate of encounter and collisions between NCOs in 
the disk. The collisional cross-section ($\sigma_{\rm coll}$) of compact NCOs of 
mass M depends on the relative velocity at infinity ($v_{\infty}$) as 
\begin{equation}
\sigma_{\rm coll} \approx \pi r_{p}(2GM/v_{\infty}^{2})
\label{eq:sigma_coll}
\end{equation}
 in the gravitational focussing regime, where $r_{p}$ is the 
separation at periastron. In AGN disks the \emph{relative} velocities 
involved in close 
encounters can be very small compared to the velocity dispersion in
star clusters (typically $\sim 50$km/s). For 
example, NCOs on circularized orbits separated in the disk by $\Delta R \sim 0.01R$
 at $R=10^{5}r_{g}$ have relative velocities due to Keplerian shear at 
periastron of only $\sim 5 \rm{km}\rm{s}^{-1}$, where $r_{g}=GM_{\rm SMBH}/c^{2}$ is the gravitational radius of the supermassive black hole of mass 
$M_{\rm SMBH}$. The disk NCO velocity dispersion varies with radius as 
$\sigma \approx \sqrt{\overline{e}^{2} + 
\overline{i}^{2}}v_{k}$, where $\overline{e},\overline{i},v_{k}$ are the mean NCO 
orbital eccentricity, mean NCO orbital inclination and Keplerian velocity 
respectively. Fig.~\ref{fig:relvel} shows the NCO velocity dispersion ($\sigma$) 
as a function of radius in a Keplerian AGN disk for a range of mean eccentricities 
and inclinations. Also shown in Fig.~\ref{fig:relvel} is the typical velocity 
dispersion in star clusters 
($\sim 50$km $\rm{s}^{-1}$, red horizontal dashed line). So $v_{\infty}$ for a 
typical interaction in a stellar cluster is $\sim 50$km $\rm{s}^{-1}$. 
Fig.~\ref{fig:relvel} shows that the velocity dispersion of NCOs at large disk 
radii is less than in star clusters for small to moderate NCO orbital 
eccentricities and inclinations ($e,i \sim 0.01-0.05$). Since the numbers of NCOs
 should increase with radius, most NCOs should live in the outer disk, where the 
NCO velocity dispersion should be smallest. 

The collisional cross-section of a seed IMBH (mass M) with compact objects (mass m) such as 
neutron stars, white dwarfs and stellar mass black holes depends on 
relative velocity as \citep{b76}
\begin{equation}
\sigma_{\rm coll}=2\pi\left(\frac{85 \pi}{6 \sqrt{2}}\right)^{2/7} \frac{G^{2}m^{2/7}M^{12/7}}{c^{10/7}v_{\infty}^{18/7}}.
\label{eq:sigma}
\end{equation} 
Numerically, this can be written as $\sigma_{\rm coll} \approx 
2\times 10^{26}m_{10}^{2/7}M_{50}^{12/7}v_{6}^{-18/7} \rm{cm}^{2}$ where 
$v_{\infty}=10^{6}v_{6} \rm{cm} \rm{s}^{-1}$ and $M_{50},m_{10}$ are in units of 
$50M_{\odot},10M_{\odot}$ respectively \citep{b81}. For 
$M_{50},m_{10}=1$, located $\sim 10^{5}r_{g}$ from a supermassive black hole, 
small eccentricities $e \sim 0.01$ in Fig.~\ref{fig:relvel} lead to collision 
cross-sections up to an order of magnitude larger than in clusters (for 
$M_{50},m_{10}=1;v_{6}=5$ above). However, for large NCO orbital eccentricities 
($e \geq 0.1$), IMBH collisions in AGN disks will have smaller cross-sections 
than in star clusters, over most of the disk. Therefore if the initial mean 
eccentricity ($\overline{e}$) of the disk NCO 
distribution is large, mechanisms for damping orbital eccentricities will be very 
important in determining whether IMBH growth via collisions is efficient in AGN 
disks.

\subsection{NCO orbital damping \& excitation}
\label{sec:damp}
\begin{figure}
\includegraphics[width=3.35in,height=3.35in]{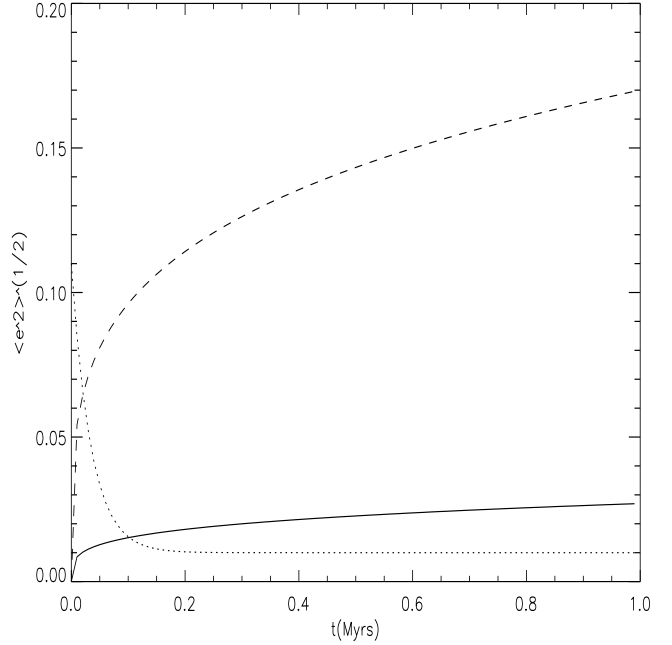}
\caption{The rms eccentricity as a function of time for an annulus of $10^{3}$ 
stars of identical mass $0.6M_{\odot}$ located in an AGN disk 
between 0.05-0.15pc (or $1-3 \times 10^{4}r_{g}$) around a $10^{8}M_{\odot}$ 
supermassive black hole. $<e^{2}>^{1/2}$ scales as $N^{1/4}, M^{1/2}$. The solid 
curve shows the relaxation of the isolated distribution of stars assuming initial circularized orbits. 
The dashed curve shows relaxation \emph{plus} dynamical heating from a population 
of $N_{i}=100N_{1}$ 
cusp stars with $M_{i}=M_{1}$ and $\overline{\sigma}_{i} \sim 0.5 v_{k}$. The 
dotted line shows the net domination of exponential damping by disk gas over 
orbital excitation by cusp stars. 
\label{fig:heating}}
\end{figure}

We begin with a fully analytic approach, demonstrating the relative importance of 
competing terms and effects. The velocity dispersion ($\sigma$) of NCOs in a disk
 is excited by dynamical heating and is damped by dynamical cooling. Thus
\begin{equation}
\frac{d \sigma}{dt}= \Delta Q_{+} -\Delta Q_{-}
\label{eq:general}
\end{equation}
where $\Delta Q_{+}$ is the dynamical heating term and $\Delta Q_{-}$ is the 
dynamical cooling term. Dynamical heating comes from two sources: the 
relaxation of disk NCOs through mutual interactions and the dynamical excitation 
of disk NCOs by cusp NCOs, so
\begin{equation}
\Delta Q_{+}= \delta Q_{\rm relax} + \delta Q_{\rm excite}.
\label{eq:heat}
\end{equation}
Considering first the relaxation term, we assume that there are $N_{1}$ stars 
of mass $M_{1}$ and velocity dispersion $\sigma_{1}$ in an annulus of width 
$\Delta R$ centered on $R$. The relaxation timescale is given by  \citep{b34}
\begin{equation}
t_{\rm relax}=\frac{2\pi C_{1}R \Delta R \sigma^{4}}{G^{2} N m^{2} {\rm ln} \Lambda \Omega}
\label{eq:trelax}
\end{equation}
so 
\begin{equation}
\delta Q_{\rm relax}=\frac{\sigma}{t_{\rm relax}}=\frac{D_{1}}{C_{1}}\frac{1}{\sigma^{3}_{1}} 
\end{equation}
where
\begin{equation}
D_{1}=\frac{G^{2}N_{1}M^{2}_{1}\rm{ln}\Lambda_{1}}{R\Delta R t_{orb}}
\end{equation}
where $\Omega$ is the Keplerian frequency, ln$\Lambda_{1}$ ($\sim 9$) is the 
Coulomb logarithm and $C_{1} \sim 2.2$. The solid curve in 
Fig.~\ref{fig:heating} shows the evolution of 
$<e^{2}>^{1/2}$ due to relaxation alone for a population of 
$N_{1}=10^{3}$ stars of mass $M_{1}=0.6M_{\odot}$ in an annulus of width 
$\Delta R=0.1$pc centered on $R=0.1$pc (equivalently $1-3\times 10^{4}r_{g}$ 
of an AGN disk around a $10^{8}M_{\odot}$ supermassive black hole). For 
these stars, $v_{k}=2100$km/s and so $t_{orb}=9\times 10^{9}$s. Since, for 
moderate eccentricities, 
\begin{equation}
\sigma=\frac{<e^{2}>^{1/2}v_{k}}{\sqrt{2}}
\end{equation}
the solid curve in Fig.~\ref{fig:heating} follows a $t^{1/4}$ form and rises 
(limited by the increase in $\sigma$ to approximately the Keplerian velocity). 
This solid curve applies to an isolated annulus of stars in the absence of 
competing effects.

Additional heating is supplied by the cusp population. The cusp stars will 
transfer kinetic energy to the 'colder' disk population and excite the 
$\sigma_{1}$ distribution of the disk 
NCOs (see \citet{b55} and Perets et al., 2012, in prep.). Following \citet{b34} $\delta Q_{\rm excite}$ has the form
\begin{equation}
\delta Q_{\rm excite}=\frac{\sigma}{t_{\rm excite}}=\frac{D_{2}}{C_{2}}\frac{\sigma_{1}}{\overline{\sigma}^{4}_{1i}}\left(1-\frac{E_{1}}{E_{i}} \right)
\end{equation}
where the cusp population $N_{i}\gg N_{1}$ has a similar mass function 
$M_{i}=M_{1}$ to the NCOs in the disk and
\begin{equation}
D_{2}=\frac{G^{2}N_{i}M_{i}M_{1}\rm{ln}\Lambda_{1i}}{R_{i}\Delta R t_{orb}}
\end{equation}
with $\overline{\sigma}_{1i}=(\sigma_{1}+\sigma_{i})/2$ and 
$E_{1,i}=3M_{1,i}\sigma^{2}_{1,i}$ is the kinetic energy. Note that 
$t_{\rm excite}$ is analagous to the $t_{\rm relax}$ 
term in equation~\ref{eq:trelax} except $N$ is now the cusp population ($N_{i}$).So $t_{\rm excite} \sim (N_{1}/N_{i})t_{\rm relax} \ll t_{\rm relax}$. We assume that on 
average $\overline{\sigma}_{i} \sim \sqrt{\overline{e}^{2}+\overline{i}^{2}}v_{k} \sim 0.5 v_{k}$ 
so $E_{i}>E_{1}$ (i.e. the cusp stars have greater kinetic energy than the disk stars). The 
dashed curve in Fig.~\ref{fig:heating} shows the addition of this excitation term to the 
evolution of $<e^{2}>^{1/2}$, assuming $N_{i}=10^{2}N_{1}$ and $C_{1}/C_{2}=3.5$ 
\citep{b34}, with ln$\Lambda_{1i}\sim$ ln$\Lambda_{1}$. Clearly, the curve retains a $t^{1/4}$ 
dependence, but at larger values of $<e^{2}>^{1/2}$. Thus, dynamical heating by stars in the cusp 
dominates relaxation by the stars in the disk \citep[see also][]{b55}.

The competing dynamical cooling term $\Delta Q_{-}$ is dominated by gas damping 
of the disk NCO orbits. Gas drag in AGN disks will tend to reduce small NCO 
orbital eccentricities and inclinations to much smaller values. Gas at co-orbital 
Lindblad 
resonances will damp (e,i) for NCOs with $q \leq 10^{-3}$ \citep[e.g.][]{b53,b43,b40,b47}. Since 
this mechanism depends on the co-rotating gas mass, both 
stellar and compact NCOs and IMBH seeds will have their orbits damped, 
particularly in the outer 
disk where most of the disk mass is located. For small eccentricities ($e<2(H/r)$), 
orbital eccentricity decays exponentially over time $\tau_{\rm e} \approx 
(H/r)^{2} \tau_{\rm mig}$, where $h=H/r$ is the disk aspect ratio and 
$\tau_{\rm mig}$ is the migration timescale \citep{b43,b47}. Thus
\begin{equation}
\frac{de}{dt}=-\kappa e
\end{equation}
which will give us a term linear in $\sigma$ as the damping term in 
$\Delta Q_{-}$. By analogy with the relaxation term ($\sigma/t_{\rm relax}$) above, we choose
 $\kappa=1/t_{\rm damp}$. The damping timescale is given by \citep[e.g.][]{b37}
\begin{equation}
t_{\rm damp}=\frac{M^{2}_{BH}h_{\rm gas}^{4}}{m\Sigma a^{2} \Omega}
\end{equation}
where $h_{\rm gas}=H/R=c_{s}/v_{k}$ is the \emph{gas} disk aspect ratio (the disk 
of stellar NCOs has an aspect ratio $h_{\rm stars}=\sigma/v_{k}$ but the disk 
NCOs are not the source of damping). For larger eccentricities ($e>2(H/r)$) the 
eccentricity damping goes as \citep{b47}
\begin{equation}
\frac{de}{dt}=-\frac{\kappa}{e^{2}}
\end{equation}
where we choose the same normalization $\kappa=1/t_{\rm damp}$ as above. So, 
\begin{equation}
\Delta Q_{-}= -\kappa\left[ \beta^{\prime}\sigma +\frac{\beta^{\prime\prime}}{\sigma^{2}}\right]
\end{equation}
where $\beta^{\prime}=1$ if $e<0.1$, zero otherwise and 
$\beta^{\prime\prime}=1$ if $e>0.1$, zero otherwise.
Thus, our expression for the combined relaxation, excitation and gas damping 
of the velocity dispersion for an annulus of NCOs is given by
\begin{equation}
\frac{d\sigma}{dt}=\frac{D_{1}}{C_{1}}\frac{1}{\sigma^{3}_{1}} + \frac{D_{2}}{C_{2}}\frac{\sigma_{1}}{\overline{\sigma}^{4}_{1i}}\left(1-\frac{E_{1}}{E_{i}} \right)-\kappa\left[ \beta^{\prime}\sigma +\frac{\beta^{\prime\prime}}{\sigma^{2}}\right]. 
\label{eq:combo}
\end{equation}
For $<e^{2}>^{1/2}<2h$ \citep{b47}, eqn.~\ref{eq:combo} has the general form
\begin{equation}
\frac{d\sigma}{dt}=\left[\frac{\sigma}{t_{\rm relax}}+\frac{\sigma}{t_{\rm excite}}\right]-\frac{\sigma}{t_{\rm damp}}=\frac{A}{\sigma^{3}} -\kappa \sigma.
\end{equation}
where we have combined the dynamical heating terms into a single general 
$A/\sigma^{3}$ term. Since $t_{\rm excite} \ll t_{\rm relax}$, at steady state, $d\sigma/dt=0$, 
and so $t_{\rm damp} \sim t_{\rm excite}$ and $e$ takes the general form
\begin{equation}
e^{4} \sim \frac{4 G^{2}N_{c} m M^{2}_{\rm BH} h_{\rm gas}^{4} \rm{ln}\Lambda}{2 \pi C_{2} \Sigma a^{2} R \Delta R v^{4}_{k}}
\end{equation}
where $N_{c}$ is the number of stars in the cusp. So, for a population of 
$10^{3}\times 0.6M_{\odot}$ stars located at $1-3 \times 10^{4}r_{g}$ in an AGN 
disk, with $h_{\rm gas}\sim 10^{-2}$ and a cusp population of $N_{c}=10^{5}$ 
stars, equilibrium eccentricity is $e \sim 0.01$. In Fig.~\ref{fig:heating} we 
plot (dotted line) the evolution of e over time assuming $i \sim 0$. From this we 
see that disk NCOs should rapidly settle down to near circular orbits 
($<e^{2}>^{1/2} \sim 0.01$) within $\sim 0.1$Myr. Therefore collision 
cross-sections ($\sigma_{\rm coll}$) of IMBH 
seeds in AGN disks should rapidly become much larger than typical collision 
cross sections in star clusters and it is gas damping that makes the difference. 

We expect gas damping to become even more dominant if the NCO disk 
population declines ($\dot{N}_{-}$) due to mergers, accretion and scatterings. 
From equipartition of energy, we expect $\dot{N}_{-}$ will be dominated by low 
mass stars. At moderate inclinations, these NCOs can be captured fairly quickly 
by the disk again (such that $\dot{N}_{+}$ increases), if the gas damps the 
orbital inclination efficiently \citep{b51}. As $\dot{N}_{+}$ increases, the 
system is driven towards a dynamical equilibrium when 
$\dot{N}_{-} \approx \dot{N}_{+}$.

One important point to note from the above discussion is that the dynamical 
heating of the NCOs by cusp stars ($\Delta Q_{+}$) gets transferred to the AGN 
disk \emph{gas}. The stability of the outskirts of the AGN disk is 
a well-known and unsolved problem \citep[e.g.][]{b22}; dynamical heating of disk 
NCOs by cusp stars is a new, additional source of disk heating which will contribute 
to maintaining the outer disk against gravitational instability. A self-consistent calculation of 
the disk heating requires a disk model \citep[e.g.][]{b22} and is beyond the 
scope of this paper, but see McKernan et al. 2012 (in prep.). Nevertheless, we 
can see that a large density of NCOs in a galactic nucleus will strongly excite 
the orbits of disk NCOs ($\delta Q_{\rm excite}$ is large). Gas damping 
($\Delta Q_{-}$) will naturally transfer much of this dynamical energy to the 
disk gas. Thus, disk luminosity must increase and the disk itself will puff up. 
The scale height increase will be a function of the density of NCOs in the 
nucleus. Therefore, one prediction of our 
model is that among nuclei with similar supermassive black hole masses, those 
with denser stellar cusps, should generate more luminous AGN disks (see Paper 
II). 

So far we have discussed low mass stars. However, we are interested in higher 
mass IMBH seeds. For simplicity let us assume a steep NCO mass function ($dN/dM 
\propto M^{-3}$) with two mass bins. The low mass 
population NCOs are $0.6M_{\odot}$ stars ($N_{l}$ in number); thus the high mass 
NCO population is $10^{-3}N_{l}$ $\times 10M_{\odot}$ stellar mass black holes. 
For a total initial disk NCO mass of $10^{3}(10^{4})M_{\odot}$, the distribution 
is $1.65 \times 10^{3}(10^{4})$ low mass stars and 1(10) stellar mass black holes.
\citet{b34} show that a low mass population of stars will diffuse out of the disk
 more than the high mass population of stars and in fact damp the orbits of the 
high mass stars, as expected from equipartition. Thus, 
for small initial values of $<e^{2}>^{1/2}$ among disk NCOs, we expect potential 
IMBH seeds in AGN disks to evolve to even smaller eccentricities than the 
equilibrium value of $e\sim 0.01$ calculated above. Recall that small 
eccentricities imply large $\sigma_{\rm coll}$, allowing IMBH seedlings to grow 
rapidly via collisions.

\subsection{Binary mergers in the disk}
\label{sec:bins}
Depending on the recent star formation history of a given galactic nucleus, massive binaries 
are likely to be rare in AGN disks. However, if there is even \emph{one} in the 
initial AGN disk, it will have the largest collisional cross-section of any disk 
NCO and should undergo the 
largest number of interactions \citep[e.g.][]{b80,b62}. A massive binary, if 
present, is therefore the most likely IMBH seed and should arise frequently 
enough to be of astrophysical interest (for similar ideas concerning 
planetesimal growth through binary-single interactions in a protoplanetary disk 
see e.g. \citet[][]{b36}). In this section, we briefly consider some of the 
issues involved in binary mergers in an AGN disk and we contrast the merger 
efficiency with that found in star clusters. For ease of comparison we consider 
the $50M_{\odot} + 10M_{\odot}$ binary from \citep{b81}.

An unequal mass binary ($M>m$, separation 
$a_{\rm bin}$) in the disk is considered hard if its binding energy ($GMm/a_{\rm
 bin}$) is greater than the kinetic energy ($m\sigma^{2}=
m(\overline{e}^{2}+\overline{i}^{2})v_{k}^{2}$) of a typical interacting NCO. The collisional 
cross-section of such a binary is given by \citep{b36}
\begin{equation}
\sigma_{\rm coll} \approx \pi a^{2}_{\rm bin} \left(\frac{v_{c}}{v_{\infty}}\right)^{2} \left(\frac{r_{\rm N}/10^{6}{\rm cm}}{a_{\rm bin}/2.14 \times 10^{8}{\rm cm}}\right)
\end{equation}
where $v_{c}=(G/\mu (Mm/a_{\rm bin}))^{1/2}$ is the critical velocity separating hard and soft 
binaries, with $\mu=(M_{\rm bin}\times M_{N})/(M_{\rm bin}+M_{\rm N})$ is the reduced total mass 
of the binary ($M_{\rm bin}$) and interacting NCO ($M_{\rm N}$). The time to merge for a binary of
 reduced mass $\mu=mM/(M+m)$ where $M> m$, semi-major axis $a_{\rm bin}$ and eccentricity 
$e_{\rm bin}$ is
\begin{equation}
\tau_{\rm merge} \approx 3 \times 10^{8}M^{3}_{\odot}(\mu M^{2})^{-1}(a_{\rm bin}/R_{\odot})^{4}(1-e^{2}_{\rm bin})^{7/2} \rm{yr}
\end{equation}
and the typical semi-major axis separation for a merger time of $\tau_{6}$Myr 
(assuming $e_{\rm bin}\sim 0$) is
\begin{equation}
a_{\rm bin} \approx 3 \times 10^{11} \tau_{6}^{1/4}M_{50}^{1/2}m_{10}^{1/4} \rm{cm}
\end{equation}
where $M_{50},m_{10}$ are the masses in units of $50M_{\odot}$ 
and $10M_{\odot}$ respectively \citep{b81}. However, the above discussion 
neglects the gas disk. 

\citet{b75} carried out hydrodynamic simulations of a binary in an AGN 
disk and found that binaries harden rapidly due to interaction with their 
own migratory spiral wakes. \citet{b75} also found that $e_{\rm bin}$ is damped 
rapidly with inward migration. The rate of binary 
hardening ($\dot{a}_{\rm bin}$) scales with the disk surface density such that 
$a_{\rm bin}/\dot{a}_{\rm bin} \ll \tau_{\rm bin}$, the binary migration 
timescale. Massive binaries with 
initial separation $a_{\rm bin} \sim 
0.3R_{H}$, end up at half this separation within 10 orbits of the supermassive 
black hole, where $R_{H}$ is the Hill radius $=(q/3)^{1/3}a$, with a the 
semi-major axis of the binary center of mass and q the mass ratio of the reduced 
mass binary to the supermassive black hole. For a constant rate 
of hardening ($\dot{a}_{\rm bin}$), a $M_{50},m_{10}$ migrating binary separated 
by $\sim 3\times 10^{11}(10^{13})$cm (or $3R_{\odot}$(2 AU)) at $10^{5}r_{g}$ 
will merge in $<0.1(2)$Myrs, which is a very small 
fraction of the AGN disk lifetime. So, because of the presence of a gas disk, 
it is easier to harden binaries in AGN disks than in star clusters.

Binaries will also encounter field NCOs in the disk as they migrate. This can 
either result in hardening to merger or disruption. The probability that
 a binary is disrupted per unit time is $1/t_{\rm dis}$ where
\begin{equation}
t_{\rm dis}=\frac{9|E|^{2}}{16 \sqrt{\pi} \nu G^{2} m^{4}\sigma}\left(1+\frac{4m\sigma^{2}}{15|E|}\right)\left[1+exp\left(\frac{3|E|}{4m\sigma^{2}}\right)\right]
\label{eq:disrupt}
\end{equation}
where the field NCO has mean mass $m$ and velocity dispersion $\sigma$, $\nu$ is 
the number density of field NCOs, $E=-GMm/a_{\rm bin}$ is the binding energy of 
the binary, and the average energy change per 
interaction is $\sim -0.2 m \sigma^{2}$ \citep{b42}. In 
stellar clusters $\sim 10^{2}$ field interactions are required to harden massive 
binaries to merger \citep{b81}, where the relative velocities at close 
encounters are approximately the velocity dispersion ($\sim 50 \rm{km} 
\rm{s}^{-1}$) in star clusters. In AGN disks, the number of interactions 
required 
to harden a binary to merger depends on the mean eccentricity ($\overline{e}$) 
of the field NCO orbits. The probability of binary disruption ($1/t_{\rm dis}$) 
increases as $\overline{e}$ increases. However, for 
already hard binaries (large $|E|$), fewer interactions (of energy 
$-0.2m\sigma^{2}$) are required for merger in gas disks. Thus, hard binaries will 
continue to harden due to inward 
Type I migration \citep{b75} \emph{and} due to interactions with field NCOs in 
the disk.

For a $60M_{\odot}$ unequal 
mass binary ($M_{50},m_{10}$) at $10^{5}r_{g}$ interacting with a population of 
NCOs having $\overline{e}=0.01$, the energy per binary interaction 
($-0.2m\sigma^{2}$) is $\sim 4\%$ the typical interaction energy in stellar 
clusters (where $\sigma \sim 50$km $\rm{s}^{-1}$). So for highly damped NCO 
orbits, binary interactions are likely to be 'soft'. The number of interactions 
depends on the disk NCO surface density. If this binary is already hard, it could
 merge within a few orbits, i.e. orders of magnitude faster than binary merger 
timescales in stellar clusters. 

\section{NCO migration \& collisions}
\label{sec:mig}
The gas in the AGN disk exerts a net torque on NCOs. This means that individual 
NCOs (and IMBH) will migrate within the gas disk, enhancing the probability of 
collision and merger. Migration is mostly inward in disks, although sometimes 
outward. This means that IMBH can migrate into new regions of the disk, in search 
of new NCO 'victims'. The IMBH feeding zone (approximately $a \pm 4R_{H}$) moves 
with the migrating IMBH. This is analagous to a giant 
planet core continuing to collide with planetesimals as it migrates through a 
protoplanetary disk \citet{b2}. Migration can stall in disks, leading to a 
pile-up (overdensity of disk NCOs). In this case rapid merger 
leading to rapid IMBH growth may occur, by analogy 
with pile-up in protoplanetary disks \citep{b37}. A detailed calculation of this 
scenario will be carried out in future work. NCOs and IMBH may also migrate
 onto the central supermassive black hole, just as protoplanets may migrate onto 
a central star. So the issue of IMBH survival in AGN disks parallels the survival 
of giant planets in protoplanetary disks.

\subsection{Type I IMBH migration}
\label{sec:type1}
NCOs with mass ratios 
$q \leq 10^{-4}$ of the mass of the central supermassive black hole will undergo 
migration in the disk analagous to Type I protoplanetary migration, on a 
timescale of 
\citep{b24}
\begin{equation}
\tau_{\rm I}=\frac{1}{N}\frac{M}{q\Sigma r^{2}} \left( \frac{H}
{r}\right)^{2}\frac{1}{\omega}
\label{eq:type1}
\end{equation}
where $M$ is the central mass, $q$ is the ratio of the satellite (NCO) mass to 
the central (supermassive black hole) mass, $\Sigma$ is the disk surface 
density, $H/r$ is the disk aspect ratio and $\omega$ is the satellite angular 
frequency. The numerical factor 
$N$ depends on the ratio of radiative to dynamical timescales and is a 
function of the power-law indices of $\Sigma, T$ and entropy \citep{b19,b24}.
Note that the Type I migration timescale decreases with increasing migrator 
mass at a given radius so more massive NCOs will migrate more quickly at 
a given disk radius. Binary NCOs face exactly the same torques and will migrate 
on the same timescale, but $q$ and $r$ in eqn.~\ref{eq:type1} are replaced with, 
the ratio of the reduced mass of the binary to the supermassive black hole and 
$a$, the location of the binary center of mass in the AGN disk respectively.

\begin{figure}
\includegraphics[width=3.35in,height=3.35in,angle=-90]{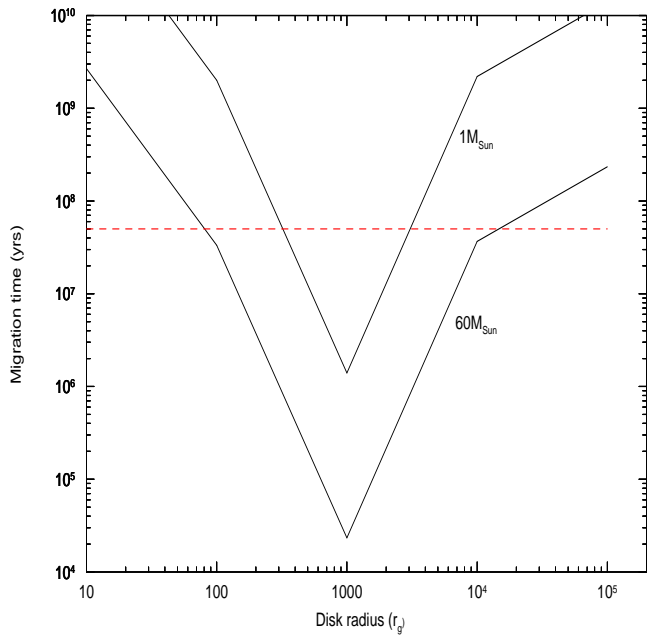}
\caption{Estimates of Type I migration timescales for $1M_{\odot}$ NCOs (upper 
curve) and $60M_{\odot}$ IMBH seedlings (lower curve) as a function of radius in 
the AGN disk modelled by \citet{b22}. Also shown 
(red dashed horizontal line) is a fiducial AGN lifetime of 50Myrs. Most disk 
NCOs should live in the outer disk 
($>10^{4}r_{g}$) for an NCO population that grows as $r^{2}$. Note that migration
 is fastest around $10^{3}r_{g}$ in the disk, where $\Sigma$ is largest and the 
disk is thinnest (H/r is smallest). There is a possibility of NCO pile-up due to 
migration at smaller disk radii, as the co-rotating mass of gas drops 
dramatically in the disk interior. 
\label{fig:migtime}}
\end{figure}

\citet{b22} model an AGN disk including all the parameters we require for 
calculation of migrator timescales as a function of radius around a 
$10^{8}M_{\odot}$ supermassive black hole. Although in principle \citet{b22} 
model a disk out to $10^{7}r_{g}$, they regard their disk as effectively 
truncating at $\sim 10^{5}r_{g}$. This 
disk also \emph{requires} a constant mass accretion rate ($\dot{M}$) and a
 constant disk viscosity ($\alpha$) at all radii
 over the disk lifetime, which are obvious simplifications. Nevertheless using 
the simple AGN disk of \citet{b22} as our disk model, we can estimate migration 
timescales semi-analytically. Figure~\ref{fig:migtime} shows the Type I migration 
timescales of a fiducial $1M_{\odot}$ NCO (upper curve) and a $60M_{\odot}$ IMBH 
seed (lower curve) as a function of disk location. The curves in Fig.~\ref{fig:migtime} 
are generated by choosing $N\sim 3$ in eqn.\ref{eq:type1} and assuming that $\Sigma$ and 
$H/r$ have the form of the curves in Fig.~2 of \citet{b22}. Also marked in 
Fig.~\ref{fig:migtime} is an approximate AGN lifetime of 50Myrs (red dashed 
line). Evidently, substantial changes of NCO orbital radius can occur even for 
low mass NCOs in the inner disk ($<10^{4}r_{g}$) over the AGN lifetime (few 
$\times 10$Myr). Larger mass migrators (stellar mass black 
holes, binaries, large mass stars or seed IMBH) are likely to have migration 
timescales roughly comparable with the AGN disk lifetime. Therefore, as IMBH 
seedlings grow in mass they should migrate in the disk and encounter low mass NCOs at
 low relative velocities. If migration stalls for a given NCO, either inwards at 
small disk radii, or outwards at large disk radii, interactions are possible at 
incredibly low relative velocities ($v_{\infty}$). Note that spiral 
density waves from migrating NCOs should not strongly perturb the NCO migrations 
or their orbits \citep{b37}. Here we assume that seed IMBH migrate independently. Of course, 
resonant capture can occur, both between IMBHs and NCOs and between multiple IMBHs, 
analagous to resonances between Jupiter and its moons, or between Neptune and Pluto. 
Given the low mass ratios and high migration speeds, an assumption of independent 
migration seems reasonable, but future simulations involving multiple migrators are 
required to test this assumption.

Because we expect NCOs in the AGN disk to migrate differentially, we expect the 
migrators to encounter each other. As large mass NCOs migrate inwards across 
the orbits of less-massive NCOs, if the gas has damped $\overline{e}$ (see 
\S\ref{sec:ecc} above) the relative velocities ($v_{\infty}$) will be low and 
the collision cross-section (with gravitational focussing) will be large. 
However, different large mass NCOs will have different outcomes from multiple 
interactions. Large stars could have a shorter life 
expectancy if there are many NCO interactions increasing the odds of merger 
and supernova. Stellar mass black holes and IMBH seedlings will 
tend to shred low mass main sequence or giant stellar NCOs as they migrate 
inwards, although they can swallow compact NCOs whole as tidal forces shred 
compact objects only after the 
compact object crosses the innermost stable circular orbit (ISCO). Inward 
migrating binaries will tend to harden and scatter lower-mass NCOs until merger 
(see below). 

Although a majority of Type I migration is directed inwards, Type I migration 
can also occur outwards in protoplanetary disks. For eccentricities $e<0.02$, 
migration may be outwards rather than inwards\citep{b47}. Migration can also 
stall on both inward and outward migrations, depending on the temperature and 
density of the adiabatic disk \citep[e.g.][]{b85,b19}. So, as we consider a 
population of NCOs migrating in an AGN disk and interacting with each other, 
we do so with the caveat that a fraction of 
the NCOs may be migrating outwards, or stalled. From 
equation~\ref{eq:type1}, the migration timescale gets longer at small disk radii 
(r), where $\Sigma$ decreases and $(H/r)$ increases due to disk heating. 
Migration may even stall or cease at small disk radii, particularly for large 
migrator masses, when the co-rotating disk mass becomes less than the migrator 
mass \citep[e.g.][]{b72,b71}. However, conditions in the hot inner disk may be 
dramatically different to the outer disk. If there is an abrupt transition to an 
optically thin accretion region, or a disk trucation or cavity, migration will 
stall. In this case, NCO pile-up can occur, potentially leading to mergers and 
ejections. Recent
 N-body simulations of protoplanetary disks suggest that migrator pile-up in 
regions of the disk where inward and outward torques balance results in very 
rapid merging \citep{b37}. If migrator pile-up occurs in AGN disks, it could
favour the rapid building of IMBH seedlings. A stalled IMBH seed can merge with 
and 
scatter piled-up NCO migrators. Using a simple equipartition of energy, an IMBH 
seedling of mass M stalled at $10^{2}r_{g}$ in an AGN disk with 
$\overline{e},\overline{i} \sim 0.01$ could scatter in-migrating NCOs (m) to 
$\sigma \sim (M/m)400$ km $\rm{s}^{-1}$. At such hypervelocities, small mass 
NCOs can be ejected into a galactic halo (see Paper II).  

If disk NCOs migrate inwards, their
 rate of migration decreases in the inner disk ($<10^{3}r_{g}$) as the disk 
surface density and co-rotating disk mass drops (see also Fig.~\ref{fig:migtime} 
above). Although conditions in the innermost AGN disk are dramatically different
from those in the outskirts, it is useful to consider the possibility of NCO 
pile-ups. Evidently, if the AGN inner disk truncates at some radius and then 
becomes a geometrically thick, optically thin advection flow, NCO migrators will 
stall near the disk truncation radius. As NCOs build up over time, the chances of 
interaction increase and hypervelocity scatterings become likely. The conditions 
may be similar to the migration trap for N protoplanets in protoplanetary disks, 
where rapid merger is possible \citep{b37}. In this case, IMBH could form in the 
inner disk with distinctive observational signatures (see Paper II for details).
 Of course, migration traps can also occur as a result of out-migration.

A majority of NCOs in galactic nuclei will have orbits that do not coincide 
with the plane of the 
AGN disk ($i \geq 0.05$) and will instead punch through the disk periodically. 
These NCOs can 
interact with each other, the disk and the migrating NCOs in the disk. NCOs on 
orbits with small radii will eventually 
decay into the plane of the disk over the AGN disk lifetime \citep{b51}, 
growing the NCO disk population ($\dot{N}_{+}$), 
particularly at small disk radii. Resonant relaxation and 
the Kozai mechanism will also allow non-disk NCOs to 
trade eccentricity with inclination and migrate into the disk over time 
\citep[e.g.][]{b88,b69,b59}. On the other hand, NCO interactions within the 
disk that lead to ejection should keep ejected NCOs at relatively low 
inclinations, thereby increasing the probability of re-capture by the disk 
(growing $\dot{N}_{+}$). A disk capture rate of $\sim 10^{-4}$ 
of the non-disk NCO population over the lifetime of the AGN disk, corresponds 
to the capture of $\sim 10^{3}M_{\odot}$ of non-disk NCOs mostly in the inner AGN disk, over 
the $\sim 50$Myr disk lifetime around a $10^{8}M_{\odot}$ supermassive black hole. 

Unlike protoplanets in disks around stars, a large number of NCOs in AGN disks should 
have retrograde orbits. The behaviour of retrograde NCOs will depend on the efficiency of
angular momentum transfer between the NCO and the disk gas. On one hand, if the coupling between NCO 
and gas is strong, the retrograde NCO rapidly loses angular momentum and falls into the central 
supermassive black hole very quickly. On the other hand, if the coupling is very weak, the disk gas
can move fast enough past the NCO that the NCO can persist in the disk for a long time without 
migrating. In this case, prograde NCOs will migrate and encounter retrograde NCOs. 

\subsection{Type II IMBH migration}
\label{sec:type2}
An IMBH that grows large enough by accreting gas can open a gap in the AGN disk 
\citep{b72}. This phenomenon is analagous to gap opening by massive planets in 
protoplanetary disks \citep{b20}. For typical disk parameters ($H/r \sim 0.05, 
\alpha=0.01$), the mass ratio required to open a gap is $q \sim 10^{-4}$. 
However, there is a strong dependency on disk viscosity for both the profile and 
depth of the gap \citep{b79,b78}. To open a gap in the disk requires low disk 
viscosity \citep{b79}
\begin{equation}
\alpha < 0.09 q^2 \left(\frac{H}{r}\right)^{-5}
\end{equation}
where $\alpha$ is the disk viscosity parameter \citep{b66}. In 
the AGN disk modelled by \citet{b22}, $H/r \sim 0.05$ on average between 
$10^{2}-10^{5}r_{g}$ and $\alpha=0.01$ is fixed. An IMBH with $q>3\times 10^{-4}$ ($>3 \times 
10^{4}M_{\odot}$) will clear a gap in this disk. The gap 
will close by pressure if $(H/r)>(q/\alpha)^{1/2}$ and by accretion if 
$(H/r)>(q^{2}/\alpha)^{1/5}$ \citep{b72}. Thus, a gap opened by a 
$3 \times 10^{4}M_{\odot}$ IMBH in the AGN disk modelled by \citet{b22} will be 
closed by pressure and/or accretion 
in the outermost and innermost parts of the disk where (H/r) and $\alpha$ are large. 
In a more viscous type of accretion flow ($\alpha \geq 0.1$, e.g. advection dominated), 
an IMBH might not open a gap in the disk. Whether an IMBH opens a gap will have 
major implications for observational signatures in AGN, but we defer that 
discussion to Paper II. 

An IMBH that opens a gap will tend to migrate on the viscous disk 
timescale (Type II migration) given by
\begin{equation}
\tau_{\rm II}=\frac{1}{\alpha}\left(\frac{h}{r}\right)^{-2}\frac{1}{\omega}. 
\label{eq:type2}
\end{equation}
From eqn.~\ref{eq:type2}, for $\alpha \sim 0.01$ and $H/r \sim 0.05$ 
(approximately the conditions across the AGN disk modelled by \citet{b22}), the 
Type II migration timescale is $\sim 10^{4} \times$ the orbital timescale. So, at 
$10^{4}(10^{5})r_{g}$, the Type II migration timescale is $\sim 1(30)$Myrs. 
Evidently a gap-opening IMBH can migrate on timescales shorter than the AGN 
lifetime across the disk. We therefore have to ask whether any gap-opening IMBH 
will survive the AGN disk? This is analagous to a major problem encountered in 
proto-planetary disk theory. The migration of some gap-opening migrators must 
somehow stall before accretion onto the central mass, otherwise no Jupiter-mass 
planets would be observed. One solution to this problem is that Type II 
migration can stall once the 
co-rotating disk mass is less than the migrator mass. This condition could arise 
due to disk drainage onto the supermassive black hole, or a change in the surface 
density profile of the disk. In the \citet{b22} disk, 
this radius is $\sim 10^{4}r_{g}$ for a $3 \times 10^{4}M_{\odot}$ IMBH. Once Type II 
migration stalls, it can resume but at a much slower rate, once the migrators' 
angular momentum is exported to the local disk \citep[e.g.][]{b72,b20}.

If we simply assume the IMBH can undergo Type II in-migration without stalling, 
the IMBH will collide with NCOs at radii interior to its starting position 
$10^{4}(10^{5})r_{g}$ before accreting onto the supermassive black hole in 
1(30Myrs). Ignoring gas accretion, the IMBH can swallow up to $5\%(50\%)$ of the 
uniformly distributed disk NCO population if it starts migrating at 
$10^{4}(10^{5}r_{g})$. So, IMBH growth via 
NCO merger can be as much as $\sim 5000M_{\odot}$/30Myr (starting at 
$10^{5}r_{g}$ and $10^{4}$ NCOs in the disk). This growth rate is due to NCO 
mergers only and does not include growth due to gas accretion (see 
\S\ref{sec:imbhgrowth} below). If the IMBH stalls permanently at 
$\sim 10^{4}r_{g}$, only a small number of remaining migrators ($\sim 2\%$ of the 
remaining disk NCO population) migrate inwards to merge with the IMBH within 
50Myrs. 

To sum up, for a powerlaw stellar mass function ($\sim M^{-3}$), 
most NCOs should be low mass stars ($<1M_{\odot}$) 
with a small fraction of compact NCOs (mostly white dwarfs, some neutron stars 
and stellar mass black holes). The largest mass NCOs (and seeds for IMBHs) 
are likely to be small in number, and consist of stellar mass black holes or 
massive binaries. IMBH seedlings will undergo Type I migration in the disk. IMBH 
migration means that they can maintain a feeding zone of disk NCOs as they 
'catch up' with the much more slowly 
migrating low mass disk NCOs. This migration of the feeding zone is precisely 
analagous to the situation expected for migrating giant planets \citep[e.g.][]{b2}. Some of the classic problems of 
protoplanetary migration (e.g. how to stop migrators from accreting onto the 
central object, or how to get migrators moving once stalled) will also apply to 
IMBH seeds in AGN disks. Nevertheless, low relative velocity encounters due to 
migration will result in large NCO collision cross-sections and will help grow 
IMBH seeds via core accretion.

\section{A model of IMBH growth in AGN disks}
\label{sec:model_growth}
In this section, we shall draw together much of the above discussion and 
construct a simple model of IMBH growth in AGN disks. The starting point for our 
model is a stellar mass black hole. This 
10$M_{\odot}$ black hole can accrete gas from the AGN disk, migrate within the 
disk and collide with disk NCOs. We follow the approach of models of giant planet
 growth in protoplanetary disks \citep{b91,b2}. In section~\ref{sec:clear} we 
discuss the growth of the IMBH seedling as a result of collision with disk NCOs 
within the IMBH feeding zone. In section~\ref{sec:imbhgrowth} we discuss the 
growth of the IMBH seedling as the result of gas accretion. 

\subsection{The parallel with 'core accretion'}
\label{sec:clear}
We considered a simple model of the growth of a $10M_{\odot}$ IMBH seed 
embedded in an AGN disk around a $10^{8}M_{\odot}$ black hole. The disk NCO 
initial population is $(10^{3})10^{4}M_{\odot}$, with a mass function of $dN/dM 
\propto M^{-3}$ (as discussed above mostly $0.6M_{\odot}$ stars). We assume that the IMBH seed 
'feeds' on NCOs within its accretion zone, given by $a\pm \delta a$ where $\delta a \sim 4R_{H}$, 
analagous to the feeding zone of giant planet cores \citep{b91}. The maximum (isolation) mass that
 the IMBH seed can attain by feeding on all the NCOs within its accretion zone is
\begin{equation}
M_{\rm ISO}=M_{I}+16\pi a^{2} \Sigma_{0} \left(\frac{q}{3}\right)^{1/3} 
\label{eq:iso}
\end{equation} 
where $\Sigma_{0}$ is the mean initial NCO surface density, $q$ is the mass ratio
 of the IMBH seed to the supermassive black hole. If there are 
$10^{4}M_{\odot}$ NCOs initially in the disk around a $10^{8}M_{\odot}$ black 
hole, the mean initial NCO surface density is $\Sigma_{0} \sim 3.5\rm{g}/\rm{cm}^{2}$. For an IMBH seed of $M_{I}=10M_{\odot}$, eqn.~\ref{eq:iso} implies 
$M_{\rm ISO} \sim 10+9(900) M_{\odot}$ at $10^{4}(10^{5})r_{g}$. So, in 
principle, a stellar mass black hole in the outer disk could grow to many times 
its original mass just by accreting low mass disk NCOs. This process is 
analagous to the growth of giant planet cores by planetesimal accretion 
\citep{b91,b20}. 

Assuming small eccentricities ($<e^{2}>^{1/2}=\Delta a/ a$) for disk NCOs, we are
 in a shear-dominated regime and the rate of mass growth of the IMBH seed may be 
approximated by the form of giant planet core growth as \citep{b20}
\begin{equation}
\frac{dM}{dt}=\frac{9}{32}\frac{(\Delta a)^{2}}{<i^{2}>^{1/2} a R_{H}}\nu \Sigma_{0} \Omega \sigma_{\rm coll}
\label{eq:coreacc}
\end{equation}
where $<i^{2}>^{1/2}$ is the rms inclination for the NCO distribution, $\nu$ is 
the relative local overdensity of disk NCOs and $\sigma_{\rm coll}$ is the 
collision cross-section as given by eqn.~\ref{eq:sigma_coll}. Thus, if 
$\Delta a= <e^{2}>^{1/2} a$ and if we assume $<e^{2}>^{1/2} \sim 2 
<i^{2}>^{1/2}$, with $v_{\infty} \sim \sigma \approx <e^{2}>^{1/2} v_{k}$, the 
rate of IMBH seed growth via core accretion within its feeding zone is 
\begin{equation}
\frac{dM}{dt}=\frac{9}{8}\frac{\nu \Sigma_{0} \Omega \pi r_{p}}{e (q/3)^{1/3}} \frac{2GM}{v^{2}_{k}}
\label{eq:coreacc_prop}
\end{equation}
where $r_{p}$ is the periastron. The periastron for compact object collisions 
with an IMBH seed is $r_{p}\ \sim R_{\odot}$ \citep{b81}. Substituting into 
eqn.~\ref{eq:coreacc_prop}, where we assume $<e^{2}>^{1/2} \sim 0.01$ is the 
equilibrium eccentricity, we find that for a $10M_{\odot}$ IMBH seed at 
$\sim 2 \times 10^{4}r_{g}$ with $\Sigma_{0}=3.5 \rm{g} \rm{cm}^{-2}$ and 
$\nu=1$, then $dM/dt \sim 10^{-7}M_{\odot}$/yr, which is approximately half the 
Eddington rate of growth. This is a very high accretion rate for a black hole, 
exceeding inferred accretion rates from gas disks in most Seyfert AGN 
\citep{b23}. Of course $\Sigma_{0}$ is the average surface density assuming a 
uniform distribution of disk NCOs and that the mass ratio of disk NCOs to gas in 
the disk is $\sim 1\%$. If, instead the mass ratio is a factor of a few larger, 
or the surface density distribution of NCOs is non-uniform, the accretion rate of
 disk NCOs by IMBH can be substantially \emph{super-Eddington}. Equally, if gas 
damping is more efficient than outlined above so that equilibrium is reached at 
$\overline{e} <0.01$ (e.g. due to a lower ratio of cusp population to disk 
NCOs), we could also reach super-Eddington rates of IMBH growth via mergers. 
From our earlier discussion, it is easy to envisage regions of the disk where 
NCOs tend to 
pile-up, leading to non-uniform distributions of disk NCOs. For example, if there
 is a region of the disk where inward and outward torques balance (as in the 
scenario outlined by \citet[][]{b37}), or migration stalls due to a change in the 
aspect ratio, or the mass of co-rotating gas drops. In these cases, we could 
write $\Sigma_{0} \sim \nu 3.5 \rm{g} \rm{cm}^{2}$, where $\nu$ is an overdensity
 factor (which could locally be $>100$ in a pile-up scenario such as in 
\citet{b37} and lead to highly super-Eddington growth). Note that a growing IMBH 
seedling avoids problems in protoplanetary coagulation theory, such as fracturing
 and sticking efficiency. For IMBH seedlings , nearby objects 
will either be captured (at high efficiency) or they will escape. 

\begin{figure}
\includegraphics[width=3.35in,height=3.35in,angle=0]{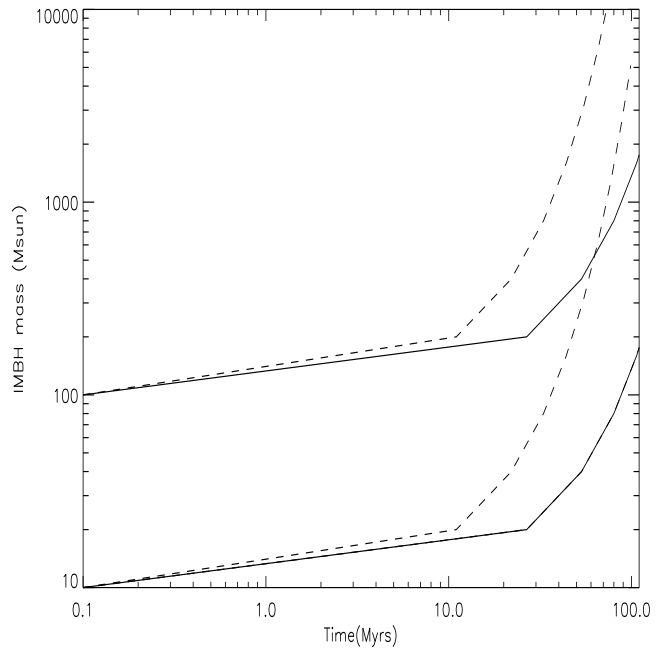}
\caption{The growth over time of a $10M_{\odot}$ IMBH seed and a $100M_{\odot}$ 
IMBH located $2\times 10^{4}r_{g}$ from a 
$10^{8}M_{\odot}$ supermassive black hole in an AGN disk. The solid (dashed) 
curves show IMBH growth as it accretes disk NCOs within $\leq 4r_{H}$ at a rate 
$\times 0.5(2.5)$ Eddington, assuming $\overline{e}=0.01$ and an initial NCO 
surface density $\Sigma_{0} \sim 3.5(18) \rm{g}\rm{cm}^{-2}$. The dashed curves 
could equivalently be generated with $\Sigma_{0} \sim 7 \rm{g}\rm{cm}^{-2}$ and 
$\overline{e}=0.004$. The gas accretion rate is assumed to be at the Eddington 
rate, for a total accretion rate of $\times 1.5(3.5)$ Eddington and a mass 
doubling time of $\sim 27(11)$Myr respectively. We assume the IMBH migrates in 
the disk and moves its feeding zone, so that it may continue to undergo 
collisions with NCOs at up to super-Eddington rates (analagous to continued 
collision and migration in protoplanetary disks \citet{b2}). In order to clear 
out $>10^{3}M_{\odot}$ of disk NCOs from this choice of initial disk location, 
\emph{outward} migration is required. Once the IMBH reaches $10^{4}M_{\odot}$ we 
assume the disk NCO population has been cleared out and the IMBH opens a gap in 
the disk, accreting gas at an Eddington rate thereafter. In this picture, between
 1/3 and 2/3 of the IMBH mass is actually due to gas accretion.
\label{fig:runaway}}
\end{figure}

Figure~\ref{fig:runaway} shows the analytic growth of a $10M_{\odot}$ stellar 
mass black hole (IMBH seed) and a $100M_{\odot}$ IMBH in an AGN disk around a 
$10^{8}M_{\odot}$ supermassive black hole. The lower curve in each case 
corresponds to $\nu=1$ (no overdensity, fiducial numbers) 
and the upper curves corresponds to a moderate over-density $\nu \sim 5$ of disk 
NCOs (or equivalently, a slight overdensity, $\nu=2$, and a moderately lower mean
 eccentricity $\overline{e}=0.004$). The fiducial mass 
doubling time for a black hole accreting gas at the Eddington rate (assuming 
$10\%$ efficiency) is $4 \times 10^{7}$ yrs. In the case of IMBH seeds growing 
via collisions in AGN disks, and assuming a gas accretion rate of $\sim$ 
Eddington, the total growth rates are $\times 1.5(3.5)$ Eddington for $\nu=1(5)$.
 At 3.5$\times$ Eddington growth rates, the mass doubling time could be as little
 as $\sim 11$Myrs. Once the IMBH reaches its isolation mass, it will then grow 
via accretion from the gas disk, at a much slower rate (Eddington or a fraction 
thereof). However, with an increased mass, the IMBH will have a shorter Type I 
migration timescale (see \S\ref{sec:mig} above). Thus, in $\sim 11$Myrs, the IMBH
 will have migrated inwards or outwards in the disk and the size of the feeding 
zone will have grown. So, a super-Eddington mode of accretion via collisions 
could continue (dashed curve in Fig.~\ref{fig:runaway}). This process is 
analagous to the migration of giant planet cores and their feeding zones in
 protoplanetary disks \citep{b2}.

\subsection{IMBH and gas accretion: runaway growth}
\label{sec:imbhgrowth}   
A big difference between IMBH growth in stellar clusters and in AGN
 disks is that in the latter, gas can damp orbits quite effectively. So mergers 
tend to be more frequent, and the IMBH seeds can continuously accrete dense gas. 
Thus, we expect IMBH growth in AGN disks to involve growth by merger (as in 
stellar clusters) but we also expect growth by gas accretion. Torques from the gas
will cause the IMBH to migrate and enter new feeding zones, analagous to the 
situation in protoplanetary disks \citep{b2}. If the IMBH grows large enough 
($q=10^{-4}$ or $10^{4}M_{\odot}$ around a $10^{8}M_{\odot}$ supermassive 
black hole), the IMBH can open a gap in the gas disk and the rate of gas 
accretion will drop. Note that although we concentrate on building a single 
IMBH, multiple IMBH seedlings ($10-100M_{\odot}$) are likely to appear in the 
disk (assuming $dM/dt \propto M^{-3}$, see discussion above).

One problem will be in preventing an IMBH from migrating onto the supermassive 
black hole. Outward migration and the stalling of migration due to a drop in 
disk surface density or a change in the disk aspect ratio are possible solutions,
 but as with protoplanetary disk theory, this theoretical problem is complicated 
and remains unsolved for now. A sufficiently massive gap-opening IMBH 
($\geq 10^{4}M_{\odot}$ around a $10^{8}M_{\odot}$ supermassive black hole) will 
grow if the AGN disk is particularly long-lived ($>50$Myrs), or if there is a 
large local disk NCO overdensity ($\nu$), or if gas damping is particularly 
efficient so that equilibrium eccentricity is $\overline{e}<0.01$. Alternatively, 
an IMBH which survives a period of AGN activity could grow to gap-opening size 
via the mechanisms outlined here, during a later, independent AGN phase. Earlier 
in the history of the Universe (at $z \sim 2$), the time between individual AGN 
phases should be much smaller \citep[e.g.][]{b93,b13,b14,b95}. So, if they 
survive, large mass (gap-opening) IMBH could grow rapidly in galactic nuclei over
 a few 100 Myrs and observational signatures of IMBH in galactic 
nuclei may be common at higher redshift (see Paper II). 

Of course, as the gas disk is 
consumed or blown away, other mechanisms will come into play. For planetesimals 
in a late-stage protoplanetary disk, planet-planet interactions, the Kozai 
mechanism or resonant relaxation can increase 
$\overline{e},\overline{i}$ \citep[e.g.][]{b88,b69,b59}. By analogy with 
protoplanetary disks, such mechanisms will certainly apply in the late stages of 
an AGN disk when most gas has been drained. However, we do not 
consider these mechanisms in more detail here since damping due to dense gas disk 
should dominate such effects (see Paper III for further discussion).

\section{Conclusions}
\label{sec:conclusions}
We show that it is possible to 
efficiently grow intermediate mass black holes (IMBH) from stars and compact 
objects within an AGN disk. Nuclear cluster objects (NCOs) in the AGN disk are 
subject to two competing effects: orbital excitation due to cusp dynamical 
heating and orbital damping due to gas in the disk. For a simple, semi-analytic 
model we show that gas damping dominates such that equilibrium eccentricities of 
disk NCOs are $e \sim 0.01$. In this case IMBH seedling formation via NCO 
collision is more efficient in the AGN disk than in stellar clusters (the 
standard model for IMBH formation). If, as we expect, gas 
damping dominates then the dynamical heating of disk NCOs by cusp stars is 
transmitted to the gas disk. This is a new, additional source of heating of the outer 
disk that can help counter the well-known gravitational instability ($Q \leq 1$) of 
the outer disk. 

Stellar mass black holes and hard massive binaries are likely IMBH seeds. IMBH 
seedlings grow by collisions with disk NCOs within their 
feeding zone ($a \pm 4R_{H}$) at near Eddington rates, as well as via gas 
accretion. IMBH seedlings will migrate within the AGN disk and so continue to 
feed on disk NCOs as they accrete gas. If there are regions of 
modest over-density of NCOs in the disk, IMBH seedling growth via collisions can 
be super-Eddington and a $10M_{\odot}$ IMBH seed 
orbiting a $10^{8}M_{\odot}$ supermassive black hole can grow to 
$\sim 300 M_{\odot}$ in less than the fiducial AGN disk lifetime. An 
over-density of disk NCOs can occur in regions of the disk where e.g. outward 
torques and inward torques balance, or where the aspect ratio changes, or where 
IMBH migration stalls. 

The largest IMBH will open gaps in AGN disks, analagous to giant planets in 
protoplanetary disks. Gap-opening IMBH are more likely to arise if: gas damping 
is very efficient (equilibrium disk NCO eccentricity is $\overline{e}<0.01$), or 
if the disk is long-lived ($>50$Myrs), or disk NCO surface density is moderately 
high ($>15$g$\rm{cm}^{-2}$), or if there is an IMBH seedling which survived a 
previous AGN phase (analagous to the survival of planets in protoplanetary 
disks). Our model of IMBH growth in AGN disks strongly parallels the growth of 
giant planets in protoplanetary disks. We leave a 
discussion of model predictions, observational constraints and implications of 
efficient IMBH growth in AGN disks to Paper II.
 
\section*{Acknowledgements}
We acknowledge very useful discussions with M. Coleman Miller on the growth of intermediate 
mass black holes and Mordecai Mac Low \& Alex Hubbard on accretion disks.


\label{lastpage}

\end{document}